\documentclass[twocolumn]{aastex62}
\usepackage{mathtools}
\usepackage[utf8]{inputenc}
\usepackage{amsmath}
\usepackage{mathtools}
\hypersetup{linkcolor=cyan,citecolor=cyan,filecolor=cyan,urlcolor=cyan}
\usepackage{color}
\definecolor{mygreen}{HTML}{139703}
\newcommand{\fid}[3]{ \left(\frac{#1}{#2} \right)^{#3} }

\defcitealias{rmp_2018}{R18}

\begin{document}
\title{How Flow Isolation May Set the Mass Scale for Super-Earth Planets}
\author{M. M. Rosenthal}
\affiliation{Department of Astronomy and Astrophysics, University of California, Santa Cruz, CA 95064, USA}
\author{R. A. Murray-Clay}
\affiliation{Department of Astronomy and Astrophysics, University of California, Santa Cruz, CA 95064, USA}

\begin{abstract}
Much recent work on planet formation has focused on the growth of planets by accretion of grains whose aerodynamic properties make them marginally coupled to the nebular gas, a theory commonly referred to as ``pebble accretion". While the rapid growth rates of pebble accretion can ameliorate some problems in planet formation theory, they raise new concerns as well. A particular issue is the preponderance of observed planets that end their growth as ``super-Earths" or ``sub-Neptunes," with masses in the range 2-10 $M_\oplus$. Once planets reach this mass scale, growth by pebble accretion is so rapid that ubiquitously ending growth at super-Earth masses is difficult unless growth rates drop at this mass scale. In this work, we highlight this issue in detail using our previously published model of pebble accretion, and also propose a reason for this change in growth rate: feedback between the growing planet's atmosphere and the gas disk inhibits accretion of smaller particle sizes by forcing them to flow around the growing planet instead of being accreted. For reasonable fiducial disk parameters this ``flow isolation" will inhibit accretion of all available particle sizes once the planet reaches super-Earth masses. We also demonstrate that the characteristics of this ``flow isolation mass" agree with previously published trends identified in the \textit{Kepler} planets.
\end{abstract}
\section{Introduction}
The \textit{Kepler} mission has provided a wealth of data about the architectures of close-in planetary systems. Chief among these results is the fact that ``Super-Earths,'' planets in the mass range between the Earth and the solar system ice giants, are extremely common in the innermost 1 au of planetary systems \citep[e.g.][]{bkb_2010,brb_2013}.  \textit{Kepler} data indicates that these planets not only far outnumber gas giants in this inner region of planetary systems---they are also an extremely common outcome of star formation, appearing around approximately one third of FGK stars (\citealt{ftc_2013}). A key question in theories of planet formation is how these planets formed, particularly due to the notable absence of any super-Earth planets in our own solar system. 

Overall trends in the \textit{Kepler} data may contain clues to the mechanisms that cause systems to preferentially form super-Earths. For example, recent analysis by \cite{wmp_2018} has shown that not only are super-Earths abundant, but within a given multi-planet system super-Earths tend to be of similar size. In addition, \cite{w_2019} has discussed the existence of a  characteristic planetary mass present in the Kepler data, which scales approximately linearly with the stellar mass $M_*$.

In this paper, we propose that these observations may be explained by the combined processes of ``pebble accretion"---rapid gas-assisted accretion of small nebular solids (e.g. \citealt{OK10}; \citealt{jl_2010}; \citealt{pmc11}; \citealt{lj12}; \citealt{OK12};  \citealt{gio_2014}; \citealt{LJ14};  \citealt{lev_num}; \citealt{mljb15}; \citealt{igm16}; \citealt{vo_2016}; \citealt{c_2016}; \citealt{jl_17}; \citealt{xbmc_2017}; \citealt{rmp_2018}; \citealt{rm_2018}; \citealt{bij_2019})---and flow isolation (\citealt{rmp_2018}, hereafter R18), a process by which coupling of these small solids to the gas flow around a planet cuts off pebble accretion at a characteristic planetary mass.  Pebble accretion requires a sufficiently massive seed to begin operating (\citealt{OK10}, \citealt{lj12}, \citealt{rmp_2018}), but once this seed mass is produced by classic planet formation processes, pebble accretion proceeds on timescales that are negligible in comparison to the evolution timescale of the gas disk (e.g. \citealt{OK10}, \citealt{lj12}, \citealt{rm_2018}), which is $\sim$Myr (e.g. \citealt{lifetimes_mamajek}). This rapid growth would naturally erase initial differences between planet masses, forcing all planets that enter this stage of accretion to halt their growth at the characteristic mass scale produced by flow isolation. Furthermore, if this mass scale is not strongly dependent on semi-major axis, this effect would lead to similarly sized planets within a given system.

The existence of a characteristic mass scale limiting planet formation is not surprising. In classical models of planet formation, planetary growth via accretion of planetesimal sized objects is initially limited by the ``isolation mass" -- the total mass in solids located inside a planet's feeding zone.  A planet grows until it has accreted all locally available material, at which point growth halts and the planet has reached its isolation mass. An isolation mass based on accretion of local solids was a key part of early theories of planet formation made to explain our solar system (see e.g. \citealt{l_1993}, \citealt{gold} for a review) and numerous works looking at a giant impacts stage of isolation mass embryos find agreement between the resultant architectures and the demographics of super-Earth systems (e.g. \citealt{hilke_iso_mass}, \citealt{dlc_2016}, \citealt{okm_2018}). 

However, the importance of an isolation mass based on local solid mass can be circumvented if pebble accretion operates, allowing planets to grow by accretion of small, mm-cm sized particles instead of $\sim$km sized ``planetesimals.'' Grains of these sizes drift radially inwards at rates much faster than the lifetime of the gas disk (e.g. \citealt{weiden_lam_drift}), ensuring that there is more mass available for accretion than just the local isolation mass. Furthermore, because they are captured on such fast timescales, accretion of these grains dominates over accretion of locally available planetesimals, allowing planets to grow far beyond their isolation mass.

However, the rapid timescales predicted by pebble accretion bring in their own challenge. Pebble accretion timescales become extremely rapid compared to the disk lifetime as planets reach terrestrial mass scales (e.g., \citetalias{rmp_2018}). If a limiting mass scale for pebble accretion is not included, these rapid growth rates imply that the final masses of planets either stall at sub-Earth masses or run away to form gas giants, with few planets finishing their growth in the super-Earth sub-Neptune mass range (\citealt{llc_2018}), which is clearly in conflict with observations of close-in planetary systems.  If pebbles are present, forming planets in this mass range thus requires some other physical process to halt growth via pebble accretion before runaway gas accretion can occur. 

Thus, both analysis of the observed \textit{Kepler} planets and theoretical considerations stemming from the efficiency of pebble accretion point to the existence of a characteristic mass scale that sets the final mass that close-in planets can reach. Several recent works (e.g. \citealt{blj_2015}, \citealt{ibr_2019}, \citealt{lmj_2019}) have looked at the architectures of systems where growth is limited by the ``pebble isolation mass," a limiting mass scale for pebble accretion first identified by \cite{ljm_2014}, which can limit growth by pebble accretion  to super-Earth masses in the inner disk. In this paper we discuss a different candidate for setting the upper mass of planets formed through pebble accretion -- the ``flow isolation mass.'' For planets growing by accreting pebbles, once planets reach a sufficient mass such that the extent of their atmosphere overtakes the impact parameter for accretion, pebbles flow around the atmosphere without being accreted, causing growth to halt. This is in contrast to the pebble isolation mass, which halts growth by raising a pressure perturbation in the gas disk, trapping pebbles exterior to the planet's orbit, as opposed to allowing them  to flow past the planet (see Section \ref{peb_iso_mass} for more discussion of the pebble isolation mass). Flow isolation naturally stops growth at terrestrial to super-Earth mass scales for reasonable fiducial disk parameters. We discuss how this mass scale emerges and is calculated, and compare predictions of the flow isolation mass with the observed population of super-Earth planets from \textit{Kepler}. In Section \ref{overview} we discuss how flow isolation operates. In Section \ref{methods} we present the details of our model, in particular how gas drag is modeled and how the impact parameter for accretion is calculated. In Section \ref{results} we present scalings and numerical results for the flow isolation mass using our fiducial disk model. In Section \ref{obs} we compare expected signatures of the flow isolation mass in the architectures of planetary system with results from the \textit{Kepler} data. Finally, in Section \ref{conclusions} we summarize our results and conclusions. 

\section{Model Overview} \label{overview}
In this section we discuss broadly how pebble accretion timescales vary as a function of mass, which leads naturally to either sub-Earth or Jupiter mass planets in the absence of a limiting mass scale. We then introduce the idea of flow isolation and explain how it modifies the planetary growth processes.

In pebble accretion, a process first reported by \cite{OK10}, \citealt{jl_2010}, and \cite{lj12}, protoplanetary cores grow by accretion of solids that are marginally coupled to the local nebular gas. These solids are both massive enough that they are not completely coupled to the gas, but not so massive that they are unaffected by gas drag. When these particles encounter growing cores, gas drag can have a substantial effect on the outcome of the interaction. In particular, gas drag can remove the relative kinetic energy between the particle and the protoplanet, gravitationally binding the particle at impact parameters where the particle would otherwise have been only deflected by the core's gravity. This increase in impact parameter can lead to dramatically faster growth rates in certain parts of parameter space. 

\begin{figure*} [htbp]
	\centering
		\includegraphics[width=7in]{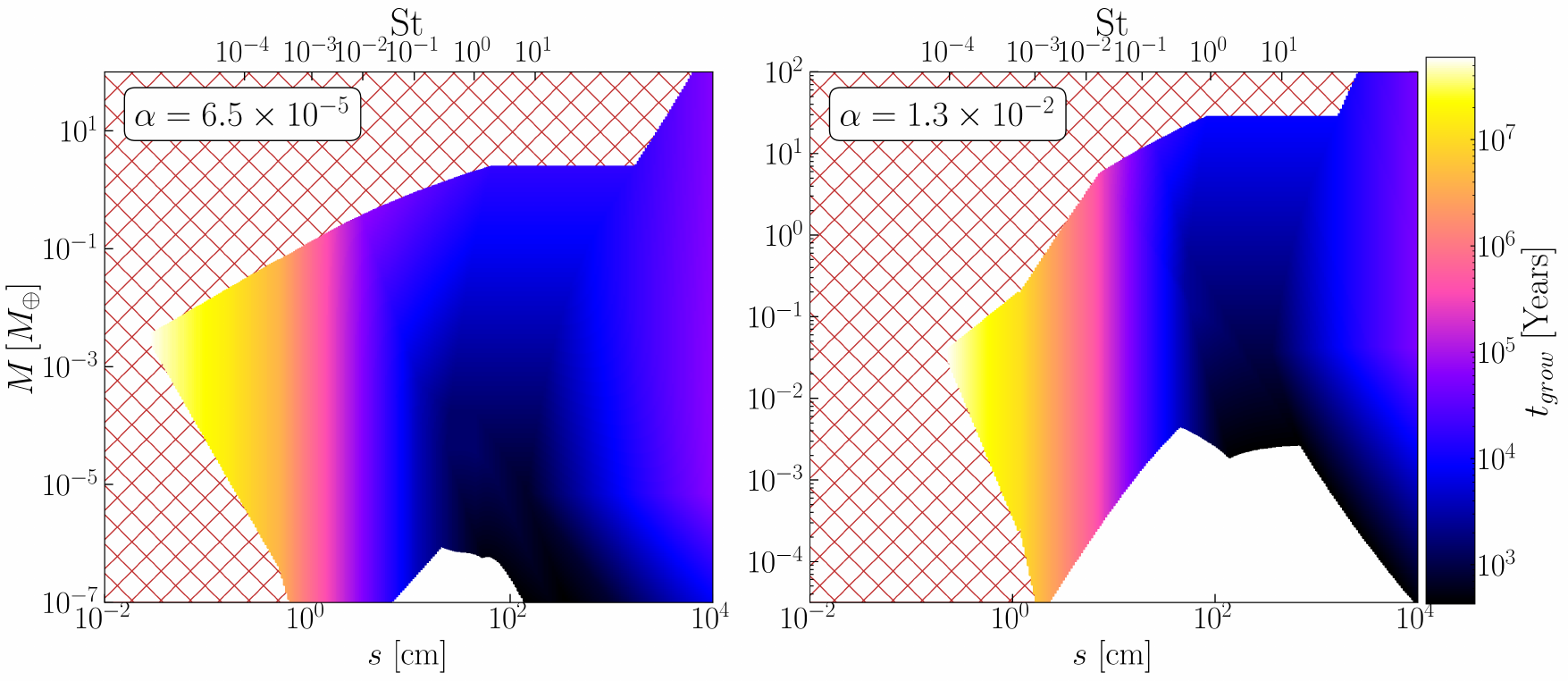}
		\caption{A plot of the growth timescale of a planet at $a=0.5 \, \rm{AU}$ undergoing pebble accretion as a function of planet mass and small body radius. The disk parameters used are described in Section \ref{fid}. The two panels show the growth timescale for two different levels of turbulence in the disk. In the lefthand side of both panels, the red hatched region indicates where growth cannot occur because pebbles flow around the core (see Section \ref{overview}
		). The white regions indicate where particles do not dissipate their kinetic energy relative to the core, and therefore cannot be accreted by pebble accretion. Pebbles in this region could still be accreted by other processes however (e.g. gravitational focusing).}
\label{fig:heatmaps}
\end{figure*}

While pebble accretion can operate at extremely fast rates, in general the timescale for growth by pebble accretion is sensitive to both the mass of the growing protoplanet and the small body size the core is accreting. An example of the pebble accretion timescale at $r=0.5 \, \rm{AU}$, using the model of \citetalias{rmp_2018}, with the disk parameters described in Section \ref{fid}, is shown in Figure \ref{fig:heatmaps}. The figure shows the growth timescale as a function of protoplanetary mass $M_\mathrm{p}$ and small body radius $s$. The two panels illustrate how growth changes in the presence of nebular turbulence, which is given in terms of the Shakura-Sunyaev $\alpha$ parameter \citep{ss_alpha}. As can be seen from Figure \ref{fig:heatmaps}, for large protoplanet masses ($M_\mathrm{p} \gtrsim 10^{-6} M_\oplus$ for the $\alpha = 6.5 \times 10^{-5}$ case, and $M_\mathrm{p} \gtrsim 10^{-3} M_\oplus$ for $\alpha = 1.3 \times 10^{-2}$) and marginally coupled particle radii ($s$ $\sim 10^{1} - 10^{3} \, \rm{cm})$, accretion occurs at an extremely rapid rate. At lower masses, however, the particle sizes that accrete on these rapid timescales are unavailable for growth, meaning the core will grow substantially more slowly.

\begin{figure}
\includegraphics[trim=0 0 180 0, clip, width=\linewidth]{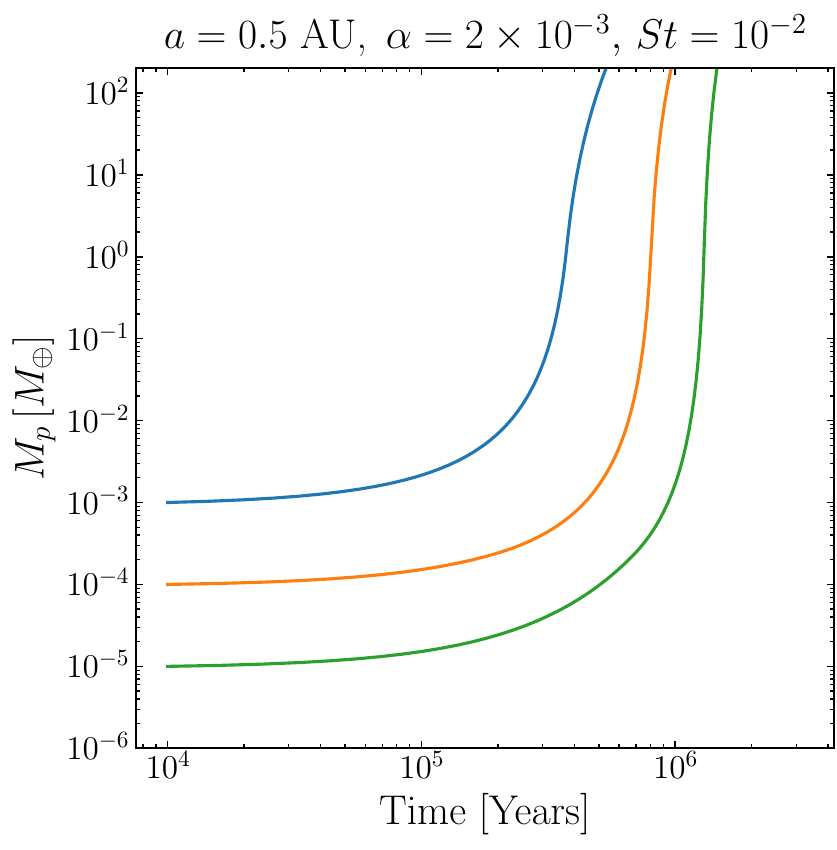}
\caption{The mass of a protoplanet undergoing pebble accretion as a function of time, for three different values of initial mass. All particles are assumed to have Stokes number of $10^{-2}$.} The disk parameters used are given in Section \ref{fid}. In all cases the protoplanet's solid mass runs away to extremely large masses on timescales shorter than the lifetime of the protoplanetary disk ($\sim$ 3 Myr).
\label{fig:t_vs_M}
\end{figure}

Because of the slower growth timescales at low core mass, these growth timescales appear to lead to binary outcomes in terms of the final planet mass. Either planets become stuck below the masses where planet formation is efficient, or they surpass this mass and grow on such rapid timescales that they easily reach $M_\mathrm{crit}$, the critical core mass needed to trigger runaway gas accretion, if growth is not halted in some manner. An example of the rapid growth timescales from pebble accretion are shown in Figure \ref{fig:t_vs_M}, which plots the mass of a protoplanet as a function of time for three different initial masses. The core grows both by gravitational focusing of pebbles (i.e. what \citealt{OK10} term the ``hyperbolic" regime) and by pebble accretion once it becomes massive enough, with the pebbles all assumed to have size $St = 10^{-2}$. Here $St$ is a dimensionless measure of particle size
\begin{align}
    St = t_\mathrm{s} \Omega
\end{align}
where $t_\mathrm{s}\equiv m v_\mathrm{rel} / F_\mathrm{D}$ is the particle's stopping time, $m$ . is the particle's mass, $v_\mathrm{rel}$ is the relative velocity between the particle and the gas, $F_\mathrm{D}$ is the drag force on the particle, and $\Omega$ is the local Keplerian angular frequency.

If the core is able to reach a mass such that $St = 10^{-2}$ particles can be captured through pebble accretion processes, growth becomes extremely fast and the planet reaches masses that are more than sufficient to trigger runaway gas accretion.
If the planet is unable to reach this point, however, planetary growth stalls at low mass. We note that in the inner regions of planetary systems, once planetesimals with $St \gg 1$ are present (see e.g. \citealt{cy10} for a review of the ``meter-size barrier"),
growth via gravitational focusing even without the assistance by gas may prevent protoplanets from stalling at masses low enough to avoid pebble accretion (see e.g. \citealt{gold}).

This discussion, however, neglects the effect of the growing planet's atmosphere on accretion. As discussed in \citetalias{rmp_2018}, as the planet grows it will accrete an atmosphere from the protoplanetary disk. Interior to the planet's atmosphere, the gas is static \footnote{Note that recent work by \citet{osk_2015} and \citet{cmk_2017} has shown that protoplanetary atmospheres may actually interact with the gas disk down to some scale, causing the atmosphere of planet's to be ``recycled". In Section \ref{atm} we give an order of magnitude calculate that demonstrates that the atmospheres of sub-thermal planets undergoing pebble accretion should be able to repel atmospheric flows at a scale comparable to $R_\mathrm{B}$}, with a density profile that rises steeply from its nebular value. Because of this, the planet's atmosphere will block the flow of nebular gas, causing the gas to flow around the planet's atmosphere (e.g. \citealt{ormel_flows}). Because of this alteration in flow pattern, particles that couple strongly to the nebular gas may flow around the core's atmosphere without being accreted. In order to determine whether particles of a given size will be diverted by the core's alteration of the gas flow, there are two criteria that must be met: 1. the maximum pebble accretion impact parameter for particles of this size must be smaller than the scale of the core's alteration of the gas flow, and 2. the time for the particle to respond to change in gas direction must be shorter than the interaction timescale between the particle and the core. 

The scale of the core's alteration of the gas flow is given by the core's Bondi radius, which is roughly the length scale at which the escape velocity from the planet is equal to the local sound speed $c_\mathrm{s}$:
\begin{align}
R_\mathrm{B} = \frac{G M_\mathrm{p}}{c_\mathrm{s}^2}
\end{align}
where $M_\mathrm{p}$ is the mass of the planet.\footnote{Note that because we are primarily interested in planets with masses less than or equal to the thermal mass---see Equation \eqref{eq:M_th}---we assume for this discussion that the planet's atmosphere is limited by the Bondi radius.}  The timescale for the particle to respond to the gas flow is the particle's stopping time, $t_\mathrm{s}$.

The maximum impact parameter at which pebble accretion could conceivably operate, $R_{\mathrm{stab}}$ is given by the scale at which gas drag balances the gravitational acceleration of the core\footnote{For sufficiently large $St$ and $M_\mathrm{p}$ we instead expect $R_\mathrm{stab} = R_\mathrm{H}$. Once $R_\mathrm{H} < R_\mathrm{B}$ the process of flow isolation is slightly modified, as the scale of the core's atmosphere is now $R_\mathrm{H}$ rather than $R_\mathrm{B}$. See \citet{rmp_2018} for a discussion of flow isolation in this regime. The full expression for $R_\mathrm{stab}$ is given in Equation \eqref{eq:r_stab}.}, that is
\begin{align} \label{eq:r_acc}
    R_{\mathrm{stab}} = \sqrt{\frac{G M_\mathrm{p} m}{F_\mathrm{D}}} \;\;.
\end{align}
Beyond this radius, even a particle that started gravitationally bound to the core would not be accreted because it would be stripped off by the gas flow.  In evaluating Equation \eqref{eq:r_acc}, $F_\mathrm{D}$ should be calculated using the relative velocity between the gas and the core at the impact parameter $R_{\mathrm{stab}}$.  This relative velocity results from either a combination of the sub-Keplerian orbital velocity of the gas and turbulent motion, which we refer to as $v_{\mathrm{gas}}$, or from Keplerian shear.

For particles to be pulled around the core by the gas, the two relevant criteria are therefore
\begin{align}
    R_{\mathrm{stab}} &< R_\mathrm{B}\\
    t_\mathrm{s} &< \frac{R_\mathrm{B}}{v_\infty} \equiv t_{\mathrm{cross}}
\end{align}
where $v_\infty$ is the velocity of the incoming particle relative to the core.
We now show that the former criterion is sufficient, as the latter is always satisfied for $R_{\mathrm{stab}} < R_\mathrm{B}$. There are two regimes for $v_\infty$: either the particle comes in with a velocity relative to the core, $v_{\mathrm{pc}} \leq v_{\mathrm{gas}}$ resulting from drift and turbulent excitation by the gas, or the Keplerian shear in the disk sets the incoming velocity, in which case $v_\infty \sim R_\mathrm{B} \Omega$, where $\Omega = \sqrt{G M_*/r^3}$ is the local Keplerian orbital frequency, $r$ is the semi-major axis of the planet, and $M_*$ is the mass of the host star. In the latter regime we have
\begin{align}
    t_{\mathrm{cross}} = \frac{R_\mathrm{B}}{\Omega R_\mathrm{B}} = \Omega^{-1}
\end{align}
and so $t_\mathrm{s} < t_{\mathrm{cross}}$ is equivalent to taking $St \equiv t_\mathrm{s} \Omega <1$, which is the regime we confine our attention to in the remainder of this work. In the former regime, we have
\begin{align}
    t_{\mathrm{cross}} = \frac{R_\mathrm{B}}{v_{\mathrm{pc}}} > \frac{R_\mathrm{B}}{v_\mathrm{{gas}}}
\end{align}
since the incoming velocity of the particle is at most the gas velocity. Rearranging Equation \eqref{eq:r_acc} and using the definition of the stopping time gives
\begin{align}
    t_\mathrm{s} = \frac{R_{\mathrm{stab}}^2 v_{\mathrm{gas}}}{G M_\mathrm{p}} = \frac{R_{\mathrm{stab}}^2}{R_\mathrm{B}^2}\frac{v_{\mathrm{gas}}^2}{c_\mathrm{s}^2} \frac{R_\mathrm{B}}{v_{\mathrm{gas}}} < \frac{R_\mathrm{B}}{v_{\mathrm{gas}}} < t_{\mathrm{cross}}
\end{align}
since $R_{\mathrm{stab}} < R_\mathrm{B}$ by assumption and $v_{\mathrm{gas}} < c_\mathrm{s}$ since all gas flows are subsonic for planetary masses less than the thermal mass (see Equations \ref{eq:v_gas} and \ref{eq:M_th}). 

\begin{figure}
\includegraphics[width=\linewidth]{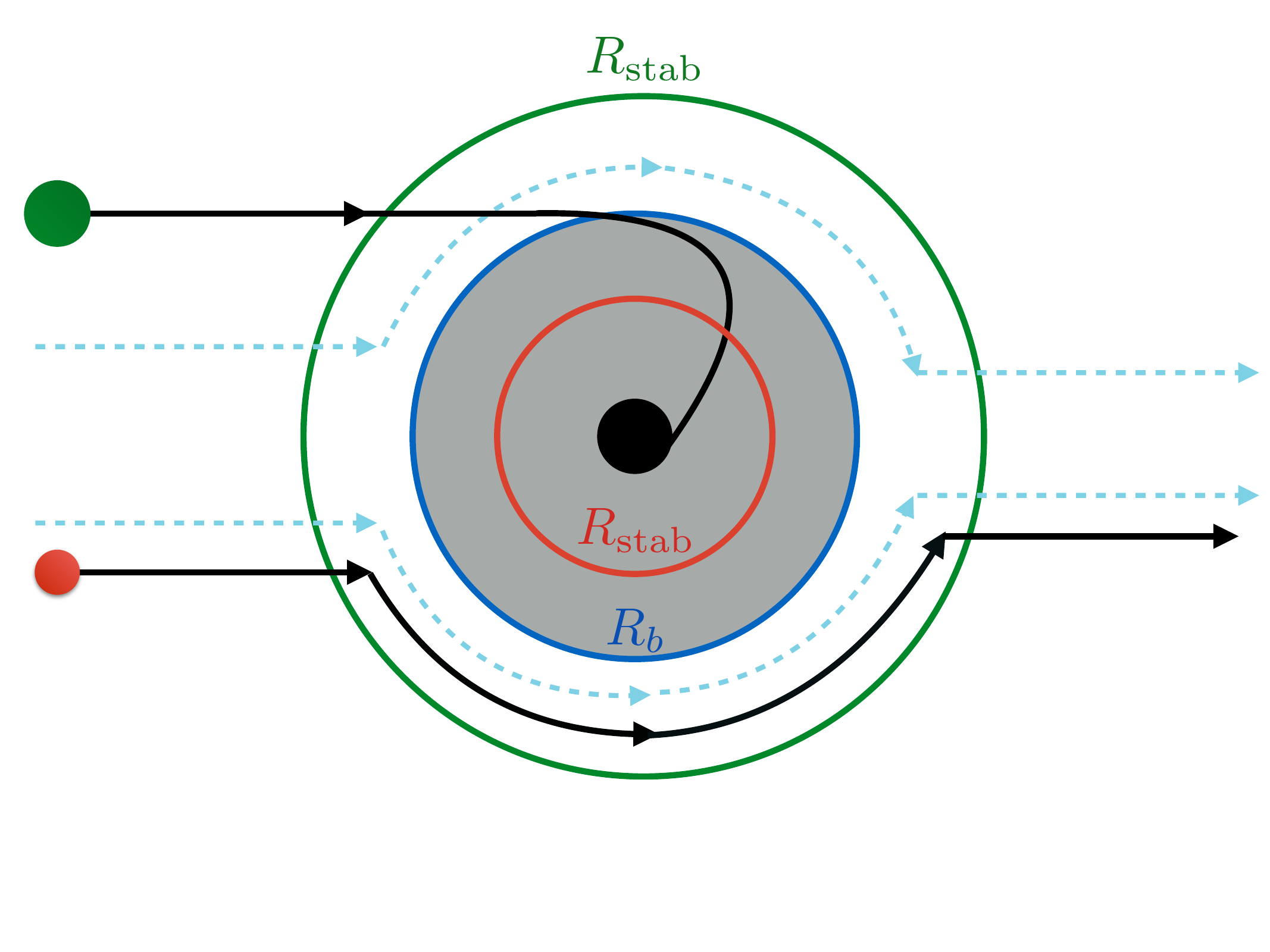}
\caption{A cartoon illustrating schematically how flow isolation operates. The planet's (black dot) atmosphere is shown by the gray shaded region, and extends up to $R_\mathrm{B}$. The gas flows around the atmosphere, as shown by the dashed blue lines. The larger, green particle, has maximal impact parameter for accretion $R_{\mathrm{stab}} > R_\mathrm{B}$, and thus can be captured at scales of $R_{\mathrm{stab}}$ before encountering the modified gas flow. The smaller red particle has $R_{\mathrm{stab}} < R_\mathrm{B}$, and is diverted by the atmosphere's modification to the flow instead of being captured.} 
\label{fig:flow_iso_illus}
\end{figure}

In summary, the only criterion that is necessary to determine whether particles will be pulled around the core's atmosphere is 
\begin{align}
R_{\mathrm{stab}} < R_\mathrm{B} \quad \text{(pebble accretion cannot operate).}
\end{align}
In pratice, this process sets the lower limit on particle sizes that can be accreted, as $R_{\mathrm{stab}}$ decreases with decreasing particle size. This process is illustrated schematically in Figure \ref{fig:flow_iso_illus}. We also note here that this cutoff in accretion is distinct from the decrease in accretion rate that occurs for smaller particle sizes, which has been discussed in other works on pebble accretion, e.g. \citet{lj12}, \citet{vo_2016}, \citetalias{rmp_2018}, and can been seen in Figure \ref{fig:heatmaps}. As an example, a 10 $M_\oplus$ core growing by accreting pebbles around a solar mass star has a growth timescale of roughly 
\begin{align}
   t_\mathrm{grow} \sim 6500 \, \mathrm{years} \fid{r}{0.5 \, \mathrm{au}}{1/2} \fid{\Sigma_\mathrm{p}}{5 \,  \mathrm{g} \, \mathrm{cm}^{-2}}{-1} St^{-2/3}
\end{align}
where $\Sigma_\mathrm{p}$ is the local pebble surface density and $r$ is the planet's semi-major axis (e.g.  \citealt{lj12}). This would require the maximal pebble size to be below $St \lesssim 10^{-4}$ for the growth timescale to exceed 3 Myr. Flow isolation, on the other hand, cuts off growth for much larger Stokes numbers; for example, in the righthand panel of Figure \ref{fig:heatmaps}, growth is shut off for all particles with $St \lesssim 10^{-1}$.

Because of the decreasing value of $R_{\mathrm{stab}}$ with decreasing particle radius, this process effectively sets a lower limit on the particle size that can be captured by pebble accretion. However, if this lower limit on particle size exceeds the maximal size of particle present in the disk, then growth of the planet will halt completely. A maximal pebble size is expected from a number of physical processes, such as a fragmentation barrier (e.g. \citealt{bke_2012}), or from radial drift in the outer disk (e.g. \citealt{bdh_2008}, \citealt{bke_2012}). For a given maximum particle size, we then have an upper limit on the mass a planet can grow to via pebble accretion, which is set by
\begin{align} \label{eq:m_flow_words}
R_{\mathrm{stab}}(St_{\rm{max}}) < R_\mathrm{B} \quad \text{(flow isolation mass)} \;.
\end{align}
In practice, we may require the impact parameter for accretion to become a factor of a few smaller than $R_\mathrm{B}$ before accretion is completely inhibited, i.e. while Equation \eqref{eq:m_flow_words} does give the scaling of the flow isolation mass, there is still some undetermined coefficient $f>1$ on the lefthand side of the equation. This constant depends on the details of the atmospheric dynamics in the vicinity of the planet, and can be determined by comparison with numerical simulations. We leave this comparison for future work. In what follows, we determine the flow isolation by determining the mass such that
\begin{align} \label{eq:m_flow_fudge}
    f R_{\mathrm{stab}}(St_{\mathrm{max}}) = R_\mathrm{B}
\end{align}
and pick $f=1.75$ for presenting our results. 

Thus Equation \eqref{eq:m_flow_fudge} defines a ``flow isolation mass," which is a function of the properties of the protoplanetary disk and the maximum particle size present (which may itself be a simple function of the disk parameters). The presence of this mass scale can halt pebble accretion at masses below the critical mass for runaway accretion of a gas envelope, allowing a super-Earth or terrestrial mass planet to remain.

\section{Methods} \label{methods}

\subsection{Fiducial Disk Model} \label{fid}
To evaluate the flow isolation mass, we use a fiducial protoplanetary disk, described by the following expressions.

Given the small semi-major axes at which super-Earths are observed, an important component of our protoplanetary disk model is viscous heating, which sets the temperature in the inner regions of protoplanetary disks. The midplane temperature from viscous heating can be determined by equating the rate of heating from accretion with radiative cooling from the midplane
\begin{align}
    \frac{G M_* \dot{M}}{r} = \frac{64 \pi}{9 }\frac{\sigma_{\mathrm{SB}} T_\mathrm{c}^4 r^2}{\tau_\mathrm{c}}
\end{align}
(e.g. \citealt{oni_2011}, \citealt{km_2011}). Here $\sigma_{\mathrm{SB}}$ is the Stefan-Boltzmann constant, $\tau_\mathrm{c}$ is the vertical  optical depth for thermal radiation escaping from the midplane, and $\dot{M}$ is the rate of mass flow through the disk. Setting $\tau_\mathrm{c} = \kappa \Sigma /2$, where $\kappa$ is the Rosseland mean opacity, the midplane temperature is given by
\begin{align}
    T_\mathrm{c} = \left[ \frac{9}{128 \pi} \frac{G M_* \dot{M} \kappa \Sigma}{\sigma_{\mathrm{SB}} r^3} \right]^{1/4}
\end{align}
If we assume a steady-state accretion disk, the disk surface density, $\Sigma$ $\dot{M}$, and $\alpha$ can be related using the equation
\begin{align} \label{eq:M_dot}
\dot{M} = 3 \pi \Sigma \nu = 3 \pi \Sigma \alpha c_\mathrm{s} H
\end{align}
where $\nu$ is the local kinematic viscosity. This gives use the freedom to fix two of $\dot{M}$, $\Sigma$, or $\alpha$; the remaining parameter can be calculated from the other two quantities using Equation \eqref{eq:M_dot}. It is common to fix $\dot{M}$ and $\alpha$, and derive the surface density from these two quantities. Doing so, however, leads to extremely large surface densities when $\alpha$ is decreased. For example, for $\alpha = 10^{-4}$, and $\dot{M} = 10^{-8} M_\odot \, \mathrm{yr}^{-1}$, the surface density at 1 au is $\Sigma \approx 12000 \, \mathrm{g} \, \mathrm{cm}^{-2}$, and the disk is Toomre $Q$ unstable for $r \gtrsim 10 \mathrm{AU}$.  Thus, in this work we choose to fix $\Sigma$ in addition to $\dot{M}$, meaning that $\alpha$ is no longer constant. 

For ease of notation, we now define the quantities
\begin{align}
    M_{*,\odot} \equiv \frac{M_*}{M_\odot}, \quad L_{*,\odot} \equiv \frac{L_*}{L_\odot}
    \end{align}
    \begin{align*}
    r_{\mathrm{AU}} = \frac{r}{\mathrm{AU}}, \quad \dot{M}_8 \equiv \frac{\dot{M}}{10^{-8} M_\odot \text{yr}^{-1}} \quad \Sigma_{3000} = \frac{\Sigma_0}{3000 \, \mathrm{g} \, \mathrm{cm}^{-2}}
\end{align*}
where $L_*$ is the stellar luminosity, and $\Sigma_0$ is the surface density at 1 au.

We choose our surface density normalization of $\Sigma_0 = 3000 \, \mathrm{g}\, \mathrm{cm}^{-2}$ from comparison with \cite{pmp_2019}  who compute disk surface densities through particle drift rates. From comparison with measured dust surface density profiles as found in e.g. \citet{andrews_09}, we also choose a power law exponent of $\gamma=1$. Thus, our fiducial surface density profile is
\begin{align}
\Sigma = 3000 \, \mathrm{g} \, \mathrm{cm}^{-2} \,  r_{\mathrm{AU}}^{-1}
\end{align}

Setting $\kappa = 0.1 \, \mathrm{cm}^2 \, \mathrm{g}^{-1}$, the fiducial temperature from viscous heating is then
\begin{align}
    T_{\text{visc}} &= 230 \, \text{K} \,  M_{*,\odot}^{1/4}\dot{M}_8^{1/4} \Sigma_{3000}^{1/4} r_{\mathrm{AU}}^{-1} \label{eq:T_visc}
\end{align}
Farther out in the disk, the disk temperature will be set by passive irradiation from the central star. We take our fiducial profile from \citet{igm16} 
\begin{align}
    T_{\text{irr}} &= 150 \, \text{K} \, M_{*,\odot}^{-1/7} L_{*,\odot}^{2/7} r_{\mathrm{AU}}^{-3/7} \label{eq:T_irr}
\end{align}
(see \citealt{cg_97} for more detail).

The temperature as a function of semi-major axis is then $T = \max(T_{\rm{visc}},T_{\rm{irr}})$, where $T_{\mathrm{visc}}$ and $T_{\mathrm{irr}}$ are given by Equation \eqref{eq:T_visc} and \eqref{eq:T_irr} respectively. The disk changes from being heated by viscous accretion to passive irradiation at a fiducial semi-major axis of
\begin{align}
    r_{\mathrm{vis-irr}} = 2.2 \, \mathrm{AU} \, \dot{M}_8^{7/16} M_{*,\odot}^{11/16} \Sigma_{3000}^{7/16} L_{*,\odot}^{-1/2}
\end{align} 
In each region, the value of $\alpha$ can be calculated using Equation \eqref{eq:M_dot}
\begin{align}
    \alpha = \begin{dcases*}
    5.3 \times 10^{-4} r_{\mathrm{AU}}^{1/2} \dot{M}_8^{3/4} M_{*,\odot}^{1/4} \Sigma_{3000}^{-5/4} \, & $r<r_\mathrm{vis-irr}$\\
    8.2 \times 10^{-4} r_{\mathrm{AU}}^{-1/14} \dot{M}_8 M_{*,\odot}^{9/14} \Sigma_{3000}^{-1} L_{*,\odot}^{-2/7} \, & $r>r_\mathrm{vis-irr}$
    \end{dcases*}
\end{align}
We also note that if global disk evolution is governed by magnetic winds, as opposed to viscous evolution, as discussed by e.g. \citet{b_2016}, then accretion heating would be reduced in the inner regions of disks. In this case, our model of a viscously heated inner disk would not be appropriate. Instead, our expressions for a passively irradiated disk would apply throughout most of the extent of the disk (as opposed to just $r>r_\mathrm{vis-irr}$), with a different regime, where the finite angular size of the star sets the irradiation, applying for $r\lesssim 0.2 \, \mathrm{au}$. See \citet{w_2019} for a discussion of this scaling.

For our fiducial disk we take the star to have solar mass, $M_* = M_\odot$, with luminosity $L_* = 3 L_\odot$, which corresponds to a solar mass star of age $\sim 1 \, \text{Myr}$ \citep{tpd_2011}. The gas has a mean molecular weight $\mu = 2.35 \, m_\mathrm{H} \approx 3.93 \times 10^{-24} \text{g}$. The neutral collision cross section in the disk is $\sigma \approx 10^{-15} \, \text{cm}^{2}$. The pebbles are taken to have density $\rho_\mathrm{s} = 2 \, \text{g} \, \text{cm}^{-3}$.

We note that the flow isolation mass is not sensitive to the solid surface density. For the calculations in this work that do require a surface density be specified (i.e Figures \ref{fig:heatmaps}, \ref{fig:t_vs_M} and \ref{fig:m_atm}), we used
\begin{align}
\Sigma_\mathrm{p} = 5 \, \text{g} \, \text{cm}^{-2} \fid{r}{\text{AU}}{-1} 
\end{align}
which is taken to match observations of the solid surface density in protoplanetary disks (\citealt{andrews_09}, \citealt{andrews}). We further note that if this surface density is converted to a mass flux using the relation $F_\mathrm{peb} = 2 \pi r v_r \Sigma_\mathrm{p}$ (e.g. \citealt{LJ14}), where $v_r \sim 2 \eta v_\mathrm{k} St$ is the radial drift velocity of pebbles, then, for the $St=10^{-2}$ particles used in producing Figure \ref{fig:t_vs_M}, this corresponds to a pebble mass flux of roughly $70  \, M_\oplus \, \mathrm{Myr}^{-1}$ in the inner, viscously heated region of the disk.

Finally, we note that we are neglecting Type I migration effects in our discussion, and instead considering expected planet masses if planets form in-situ at their observed locations.

We now quantitatively discuss how to calculate the mass scale where flow isolation occurs. We also discuss the properties of the atmospheres of cores undergoing pebble accretion. 

\subsection{Summary of Pebble Accretion Model}
In Section \ref{gas_drag} and \ref{impact_parameter} we briefly summarize how the maximum impact parameter for pebble accretion, $R_{\mathrm{stab}}$ is calculated in the model of \citet{rmp_2018}; see \citetalias{rmp_2018} for more detail.

\subsubsection{Stokes number and Gas Drag Regimes} \label{gas_drag}
The relevant parameter for measuring particle size in pebble accretion is the particle's Stokes number, $St$. The Stokes number measures particle size in terms of how well coupled the particle is to the gas, and is given by
\begin{align}
St \equiv t_\mathrm{s} \Omega \; . 
\end{align}
Here $t_\mathrm{s}$ is the particle's stopping time, and $\Omega$ is the local Keplerian angular frequency. Particles with $St\sim 1$ are maximally affected by gas drag, while particles with $St \ll 1$ are strongly coupled to the gas, and particles with $St \gg 1$ are decoupled from the gas flow. Calculation of the particle radius $s$ for which $St \sim 1$ yields radii in the eponymous ``pebble'' size range of mm-cm, particularly in the outer disk. 

Thus, in order to calculate the particle's Stokes number we first need to determine the drag force on the particle. The gas drag force on the pebbles is split into two regimes -- a ``diffuse regime,'' which applies for $s < 9 \lambda / 4$, and a ``fluid regime,'' which holds for $s > 9 \lambda / 4$. Here $s$ is the radius of the pebble, $\lambda = \mu/(\rho_\mathrm{g} \sigma)$ is the mean free path of the gas molecules, $\rho_\mathrm{g} = H/(2 \Sigma)$ is the volumetric mass density of the gas, and $H = c_\mathrm{s}/\Omega$ is the scale height of the gas disk.
The particle is in the fluid regime for
\begin{align}
    St \gtrsim \begin{dcases*}
    3.4 \times 10^{-3} r_{\mathrm{AU}}^3 M_8^{1/8} M_{*,\odot}^{-3/8} \Sigma_{3000}^{-15/8} \, & $r<r_\mathrm{vis-irr}$\\
    2.8 \times 10^{-3} r_{\mathrm{AU}}^{23/7} L_{*,\odot}^{2/7} M_{*,\odot}^{-4/7} \Sigma_{3000}^{-2} \, & $r>r_\mathrm{vis-irr}$
    \end{dcases*}
\end{align}
In the diffuse regime, the drag force is given by the Epstein drag law
\begin{align} \label{eq:epstein}
F_{\mathrm{D},\mathrm{eps}} = \frac{4}{3}\pi\rho_\mathrm{g}v_\mathrm{th} v_{\rm{rel}} s^2 \; ,
\end{align}
where $v_\mathrm{th} = \sqrt{8/\pi} c_\mathrm{s}$ is the average thermal velocity of the gas particles, and $v_{\rm{rel}}$ is the relative velocity between the particle and the gas. Assuming spherically symmetric particles of uniform density $\rho_\mathrm{s}$, the stopping time of a particle in the Epstein regime is
\begin{align}
t_{s,\rm{Eps}} = \frac{\rho_\mathrm{s}}{\rho_\mathrm{g}}\frac{s}{v_\mathrm{th}}
\end{align}
which is independent of the small body's velocity. 

In the fluid regime, the drag force depends on the Reynolds number of the particle, $Re = 2 s v_{\rm{rel}} / \left(0.5 \, v_\mathrm{th} \lambda \right)$, and can be approximated by
\begin{align}
    F_\mathrm{D} = \begin{dcases*}
    3 \pi \rho_\mathrm{g} v_\mathrm{th} v_{\text{rel}} \lambda s \quad Re < 1, \, \mathrm{Stokes}\\
    0.22 \pi \rho_\mathrm{g} v_{\text{rel}}^2 s^2 \quad Re \gtrsim 800, \, \mathrm{Ram}
    \end{dcases*}
\end{align}
Note that the Stokes regime is a linear drag regime, and the stopping time of a particle in the Stokes regime is given by
\begin{align}
    t_{s,\mathrm{Stokes}} = \frac{4}{9} \frac{\rho_\mathrm{s}}{\rho_\mathrm{g}} \frac{s^2}{v_\mathrm{th} \lambda}
\end{align}
Generally a smoothing function is employed to transition cleanly between the Stokes and Ram regimes (e.g. \citealt{cheng_drag}). In order to make the effect of various drag regimes clear in our results, we instead choose to use a piecewise drag function that transitions between the Stokes and Ram regimes at the Reynolds number for which the drag forces are equal. That is, we take the drag force in the fluid regime to be given by
\begin{align} \label{eq:f_d_piece}
    F_\mathrm{D} = \begin{dcases*}
    \frac{3 \pi}{4} \rho_\mathrm{g} v_\mathrm{th}^2 \lambda^2 Re \quad Re \leq \frac{12}{0.22}, \, \mathrm{Stokes}\\
    \frac{0.22 \pi}{16} \rho_\mathrm{g} v_\mathrm{th}^2 \lambda^2 Re^2 \quad Re > \, \frac{12}{0.22}, \, \mathrm{Ram}
    \end{dcases*}
\end{align}
This slightly underestimates the drag force on the particle at intermediate Reynolds numbers, which increases the calculated impact parameter for accretion (see Equation \ref{eq:R_WS_full}) and therefore slightly increases the flow isolation mass, as the core must get to larger masses before the Bondi radius exceeds the impact parameter for accretion.

In the ram regime, the stopping time is dependent on velocity, meaning that, for a given particle size $s$, $t_\mathrm{s}$ must be solved for numerically, using $v_{\rm{rel}}(t_\mathrm{s})$. The relevant equations for the laminar and turbulent components of the relative velocity between the particle and the gas respectively are

\begin{align} \label{eq:v_pg_lam}
v_{\mathrm{pg},\ell} = \eta v_\mathrm{k} St \frac{ \sqrt{4+St^2} }{1+St^2} \; 
\end{align}
\citep{nag}, and

\begin{align} \label{eq:RMSturb_gas}
v_{\mathrm{pg},t}^2=v_{\text{gas},t}^2\left(\frac{St^2(1-Re_\mathrm{t}^{-\frac{1}{2}})}{(St+1)(St+Re_\mathrm{t}^{-\frac{1}{2}})}\right) \; 
\end{align}
\citep{oc07}. Here $\eta \equiv c_\mathrm{s}^2/\left(2 v_\mathrm{k} \right)$ is a measure of pressure support in the gas disk, $v_\mathrm{k} = r \Omega$ is the local Keplerian orbital velocity, and $Re_\mathrm{t} \equiv \alpha c_\mathrm{s} H / (v_\mathrm{th} \lambda )$ is the Reynolds number of the turbulence, given in terms of the  Shakura-Sunyaev $\alpha$ parameter, which we use to parameterize the strength of turbulence in the disk.  In terms of $\alpha$, the root-mean-square (RMS) turbulent gas velocity is given by
\begin{align} \label{eq:ss_alpha}
v_{\rm{gas},t} = \sqrt{\alpha} c_\mathrm{s} \; .
\end{align}
Finally, the total RMS velocity between the particle and the gas is given by
\begin{align} \label{eq:v_pg}
v_{\mathrm{pg}} = \sqrt{v_{\mathrm{pg},\ell}^2 + v_{\mathrm{pg},t}^2} \; .
\end{align}
\subsubsection{Calculation of Impact Parameter for Pebble Accretion} \label{impact_parameter}
Flow isolation occurs when the impact parameter for accretion, $R_{\mathrm{stab}}$, shrinks below the core's Bondi radius. In this section, we discuss in detail how $R_{\mathrm{stab}}$ is calculated.

The scale at which a growing planet's gravity dominates over the stellar gravity is the planet's Hill radius, which is given by \begin{align}
R_\mathrm{H} = r \left( \frac{M_\mathrm{p}}{3 M_*} \right)^{1/3}
\;,
\end{align}
and $M_\mathrm{p}$ is the mass of the planet \citep{hill}. In the most favorable cases, pebble accretion allows cores to accrete over the entirety of their Hill radii (e.g. \citealt{lj12}, \citetalias{rmp_2018}), resulting in extremely rapid growth timescales relative to gravitational focusing of planetesimals.\footnote{Accretion at $R_\mathrm{H}$ is faster than gravitationally focusing a population of small bodies with velocity dispersion $v_H \equiv R_\mathrm{H} \Omega$ (which leads to the fastest growth rate in the absence of a mechanism to damp planetesimal velocities) by a factor of $R_\mathrm{H}/R_p\sim r/R_*$, where $R_p$ is the planet's radius, $r$ is the semi-major axis of the planet, and $R_*$ is the stellar radius.}
However, in order for pebble accretion to operate, the core's gravitational force needs to dominate over the force on the particle due to gas drag, in addition to the stellar tidal gravity (e.g. \citealt{pmc11}). Balancing the core's gravity with the differential acceleration due to gas drag leads to a scale is known as the wind-shearing (WISH) radius, which is given by
\begin{align} \label{eq:R_WS_full}
R_\mathrm{WS}^\prime = \sqrt{\frac{G \left(M_\mathrm{p} + m\right)}{\Delta a_\mathrm{WS}}} \approx \sqrt{\frac{G M_\mathrm{p} t_\mathrm{s}}{v_{\rm{rel}}}}
\end{align}
\citep{pmc11}. Here $m$ is the mass of the small body, $\Delta a_\mathrm{WS}$ is the relative acceleration between the protoplanet and the small body due to gas drag, and $v_{\rm{rel}}$ is the relative velocity between the small body and the nebular gas. In the second equality we've assumed that $M_\mathrm{p} \gg m$. 

In order to calculate $R_\mathrm{WS}^\prime$, we need to determine the relevant velocity for determining the drag force. As the particle approaches the core, the particle will be slowed relative to the gas flow, increasing the drag force it feels. In the most restrictive case, the particle will feel the full velocity of the gas relative to the core, which is assumed to be massive enough that it moves at the local Keplerian orbital velocity. The local gas velocity is a combination  of two factors: motion of the gas relative to the Keplerian velocity, and shear in the disk. 

The motion of the gas relative to the local Keplerian velocity has both a laminar component and a turbulent component. The laminar component arises from pressure support in the disk, which causes the gas disk to rotate at a slightly sub-Keplerian orbital velocity
\begin{align}
v_{\rm{gas,lam}} = \frac{c_\mathrm{s}^2}{2 v_\mathrm{k}} = \eta v_\mathrm{k}
\end{align}
As discussed previously, the amount of turbulence in the disk is parameterized by the Shakura-Sunyaev $\alpha$ parameter (see Equation \ref{eq:ss_alpha}). The total RMS velocity of the gas relative to the local Keplerian velocity is
\begin{align} \label{eq:v_gas}
v_{\rm{gas}} = \sqrt{\eta^2 v_\mathrm{k}^2 + \alpha c_\mathrm{s}^2}
\end{align}
(e.g. \citetalias{rmp_2018}).

The second factor contributing to the relative velocity between the gas and the local Keplerian velocity is shear in the disk. Because orbital velocity decreases as we move outwards in the disk, particles separated in the radial direction move relative to one another in the azimuthal direction. This shear velocity is of order
\begin{align} \label{eq:v_shear}
v_{\rm{shear}} = R \Omega
\end{align}
where $R$ is the separation between the particles. 

If we set $v_{\rm{rel}} = \max(v_{\rm{gas}},v_{\rm{shear}})$, then we have two measures of the impact parameter for accretion. In the former case, where $v_{\rm{rel}}=v_{\rm{gas}}$, we refer to the impact parameter as $R_\mathrm{WS}$ (i.e. unprimed); in the latter case we refer to the impact parameter as $R_{\rm{shear}}$. For a particle in a linear drag regime, there are simple analytic forms for $R_\mathrm{WS}$ and $R_{shear}$:
\begin{align}
R_\mathrm{WS} &= R_\mathrm{H} \sqrt{3 St \left( \frac{v_H}{v_{\rm{gas}}} \right)} \label{eq:R_WS_lin}\\
R_{\rm{shear}} &= R_\mathrm{H} \left(3 St\right)^{1/3} \label{eq:R_shear_lin}
\end{align}
For a particle in a nonlinear drag regime, the values of these parameters are calculated numerically. See \citetalias{rmp_2018} for a comparison of this method of modeling of impact parameter with other works. In general, the impact parameter for accretion is given by
\begin{align} \label{eq:r_stab}
R_{\mathrm{stab}} = \min\left(R_\mathrm{WS},R_{\rm{shear}},R_\mathrm{H}\right)
\end{align}

\subsection{Calculation of the Flow Isolation Mass}
\subsubsection{Analytic Calculation for Linear Drag Regimes}
As can be seen from Equations \eqref{eq:R_WS_lin} and \eqref{eq:R_shear_lin}, the impact parameter for pebble accretion decreases as small body radius is decreased. Thus, the requirement that pebble accretion can only operate for $f R_{\mathrm{stab}} > R_\mathrm{B}$ translates into an lower limit on the small body radius that can captured via pebble accretion. In a linear drag regime, where a particle's Stokes number is  independent of velocity and depends only upon particle and disk properties, we can substitute equations \eqref{eq:R_WS_lin} and \eqref{eq:R_shear_lin} into Equation \eqref{eq:m_flow_fudge} and solve for $St$. Doing so yields
\begin{multline} \label{eq:st_low}
St_{\rm{min}} = \mathrm{max} \left[f^{-2} \left(\frac{H}{r}\right)^{-3} \left(\frac{v_{\rm{gas}}}{c_\mathrm{s}}\right) \left(\frac{M_\mathrm{p}}{M_*}\right), \right. \\ \left. f^{-3} \left( \frac{H}{r} \right)^{-6} \left( \frac{M_\mathrm{p}}{M_*} \right)^2 \right] \; .
\end{multline}
where $f$ is the undetermined coefficient introduced in Equation \eqref{eq:m_flow_fudge}. Thus, if particles only exist up to some maximum size $St_{\rm{max}}$, then we can translate Equation \eqref{eq:st_low} to an upper limit on planet mass
\begin{align} \label{eq:m_flow_analy}
\frac{M_{\mathrm{flow}}}{M_*} = \min \left[f^2 \frac{c_\mathrm{s}}{v_{\rm{gas}} } \left(\frac{H}{r}\right)^3  St_{\rm{max}}, f^{3/2} \left(\frac{H}{r}\right)^3 \sqrt{St_{\rm{max}}} \right]
\end{align}
We note again that this analytic expression is only valid if the particle is in a linear drag regime; the general numerical procedure for calculating $M_\mathrm{flow}$ is discussed in the next section. 
Once the core grows to a mass such that $R_\mathrm{B} > R_\mathrm{H}$, the core's atmosphere will begin to be limited by tidal effects. In this regime the extent of the core's atmosphere, $R_\mathrm{atm}$ will now extend to $R_\mathrm{H}$ as opposed to $R_\mathrm{B}$, and we will have $R_{\mathrm{stab}} \leq R_\mathrm{atm}$ regardless of small body size. In this regime, \citet{rmp_2018} argue that growth by pebble accretion is completely halted. Given the order of magnitude nature of this argument, we again introduce an order unity factor when solving for the mass scale, which should be calibrated from numerical simulations. Because the physical processes important in this regime differ from those that dominate at lower masses (for example, the velocity difference between the planetary atmosphere and the background gas becomes supersonic), we use a different order unity constant, $f^\prime$, when determining this mass scale. Solving $R_\mathrm{B} = f^\prime R_\mathrm{H}$ for planet mass gives
\begin{align} \label{eq:M_th}
\frac{M_{\mathrm{p},\rm{max}}}{M_*} = \left( \frac{f^{\prime 3}}{3} \right)^{1/2}\frac{c_\mathrm{s}^3}{G \Omega} =  \left( \frac{f^{\prime 3}}{3} \right)^{1/2}\left(H/r\right)^3
\end{align}
This is similar in scale to the thermal mass, an often cited scale at which a growing planet is able to open a gap in the gas disk (\citealt{lin_gap}). At the thermal mass, $R_\mathrm{H} \sim R_\mathrm{B} \sim H$, though the exact form of the expression for the thermal mass depends on which of these two length scales are set equal. For the purposes of this work we define the thermal mass as the scale at which $R_\mathrm{B} = H$, in which case the thermal mass is given by 
\begin{align}
    M_{\mathrm{th}} = 3 \left(\frac{H}{r}\right)^3 M_*
\end{align}
Note that while we use this definition of thermal mass when expressing our results in terms of $M_\mathrm{p}/M_\mathrm{th}$, this definition of thermal mass makes no difference in the calculated value of the flow isolation mass, which is more fundamentally given by Equation \eqref{eq:m_flow_analy}. A different definition of the thermal mass would simply introduce additional prefactors into equations such as \eqref{eq:m_flow_th}.

In terms of the thermal mass, we can write the full expression for the flow isolation mass as
\begin{align} \label{eq:m_flow_th}
    \frac{M_{\rm{flow}}}{M_{\rm{th}}} = \min \left[f^2 \frac{c_\mathrm{s}}{3 v_{\rm{gas}}} St_{\rm{max}}, \frac{f^{3/2}}{3} \sqrt{St_{\rm{max}}}, \left(\frac{f^\prime}{3}\right)^{3/2} \right]
\end{align}
To maintain simplicity in presenting our results we set $f^\prime = f = 1.75$ in what follows.

\citetalias{rmp_2018} previously used the term ``Flow Isolation Mass" to refer to scenario where $R_\mathrm{B} > R_\mathrm{H}$, indicating that pebbles of all sizes were inhibited from accreting. However, if pebbles exist up to some maximum size, then growth can halt because pebbles of the maximal size are inhibited from accreting from the constraint in Equation \eqref{eq:st_low}. This limits planetary growth to masses lower than the thermal mass. In this work we expand the term ``Flow Isolation Mass" to include this case as well.

\subsubsection{General Numerical Procedure} \label{sec:num}
In this section we sketch the general procedure to calculate $M_\mathrm{flow}$ numerically.

If the particle is not in a linear drag regime then $St$ can no longer be defined without reference to the relative velocity between the particle and gas. In this work, we define the particle's Stokes number in a non-linear drag regime with respect to $v_\mathrm{pg}$ as defined by Equation \eqref{eq:v_pg}. Thus for a given maximum Stokes number, the algorithm to calculate $M_\mathrm{flow}$ is as follows

\begin{enumerate}
    \item Use $St_\mathrm{max}$ to calculate $v_{\mathrm{pg}}$, using Equations \eqref{eq:v_pg_lam}--\eqref{eq:v_pg}.
    \item Use the calculated value of $v_\mathrm{pg}$ to solve the equation $F_\mathrm{D} = m v_\mathrm{pg} / t_\mathrm{s}$ for particle size $s$, using Equation \eqref{eq:f_d_piece} to relate $F_\mathrm{D}$ and $s$.
    \item \label{step3} Solve for the masses such that $R_\mathrm{B} = f R_{\mathrm{WS}}^\prime$, where $R_{\mathrm{WS}}^\prime$ refers to the two solutions to the equation $F_\mathrm{D} = G M_\mathrm{p}/ R_{\mathrm{WS}}^{\prime 2}$ (Equation \ref{eq:R_WS_full}), when the drag force is calculated using a) $v_{\mathrm{gas}} = \sqrt{\eta^2 v_\mathrm{k}^2 + \alpha c_\mathrm{s}^2}$ (Equation \ref{eq:v_gas}) and b) $v_\mathrm{shear} = R_\mathrm{WS}^\prime \Omega$ (Equation \ref{eq:v_shear}). Note that in the latter case the velocity, and therefore the drag force, is also function of impact parameter.
    \item Finally, the flow isolation mass is the minimum of three mass scales: the two masses calculated in \ref{step3}.~above, and the mass scale where $f^\prime R_\mathrm{H} = R_\mathrm{B}$ defined in Equation \eqref{eq:M_th}.
\end{enumerate}
We remind the reader that we use $f^\prime = f = 1.75$ for presenting our results. Note that several, if not all, of the solutions described above need to be performed numerically, particularly if a more complicated drag law is used (e.g. the previously discussed \citealt{cheng_drag} smoothed drag law) instead of our simpler, piecewise prescription.

\subsection{Structure of Planetary Atmospheres} \label{atm}

\begin{figure} [htbp]
	\centering
		\includegraphics[width=\linewidth]{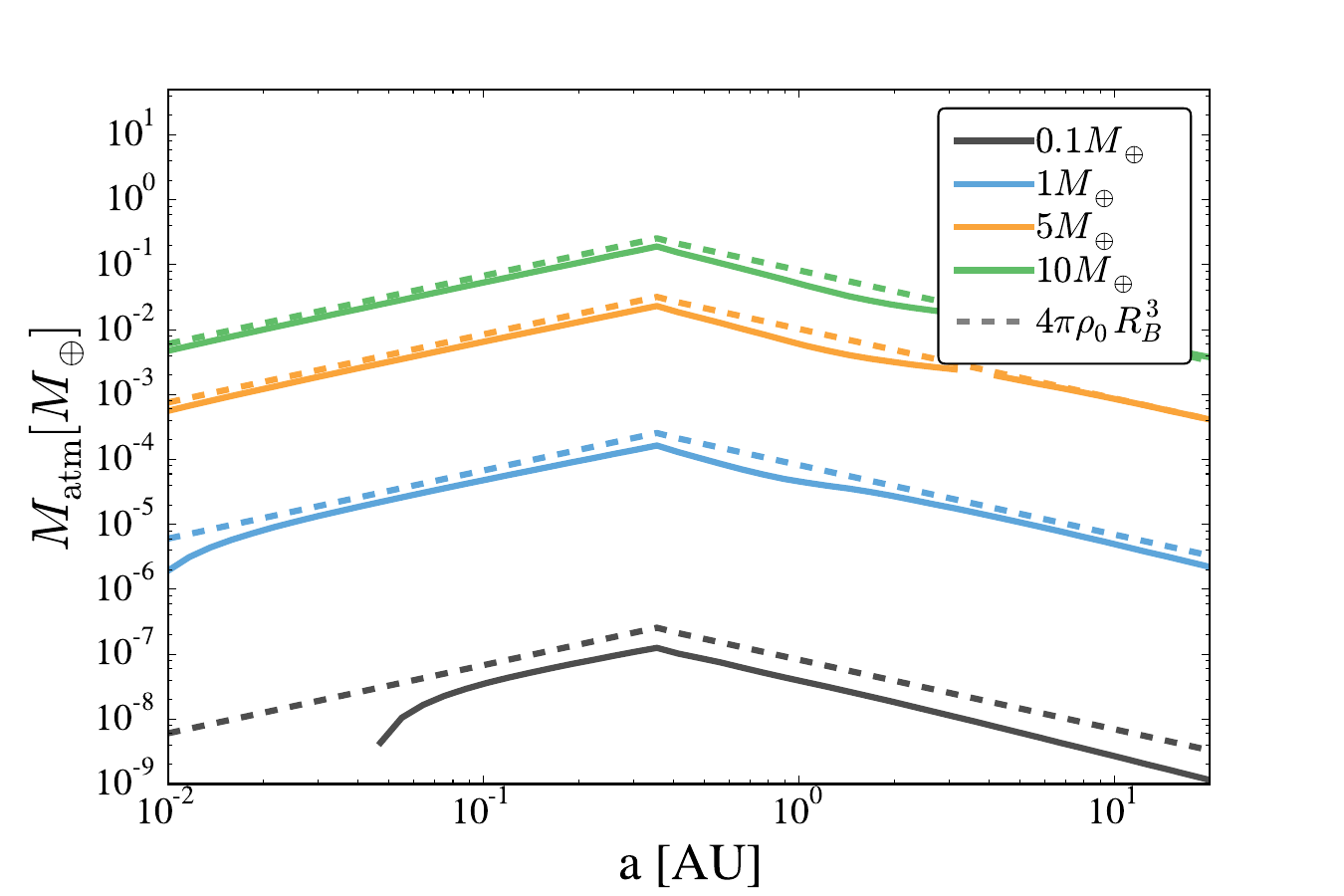}
        \epsscale{1.2}
		\caption{The atmospheric mass of a planet accreting at the maximal pebble accretion rate as a function of semi-major axis, using mixing length theory to calculate the temperature gradient. While the atmospheric mass is slightly reduced from the
		fully convective value, the decrease is relatively modest.}
\label{fig:m_atm}
\end{figure}

Equation \eqref{eq:m_flow_th} is the key result of our paper.  In deriving this expression, we have assumed that the atmosphere of the growing core is able to repel the flow of nebular gas.  In this section, we discuss the atmospheric properties of planets undergoing pebble accretion, in particular to ensure that the mass of the atmosphere is still substantial enough to act as an effective obstacle.

As planets approach the flow isolation mass, pebble accretion rates are generally extremely rapid (see Figure \ref{fig:heatmaps}). At these masses, a large fraction of the available pebble sizes will be accreted over the extent of the planet's Hill radius (e.g. \citealt{OK10}, \citealt{lj12}). This leads to a growth timescale that is independent of small body radius
\begin{align}
t_{\rm{Hill}} = \frac{M_\mathrm{p}}{2 \Sigma_\mathrm{p} R_\mathrm{H}^2 \Omega} \sim 4 \times 10^{3} \, \text{years} \left( \frac{r}{\text{AU}}\right)^{1/2} \left( \frac{M_\mathrm{p}}{M_\oplus} \right)^{1/3}  \; .
\end{align}
where $\Sigma_\mathrm{p}$ is the pebble surface density, and we have used our fiducial disk model in the second expression (see Section \ref{fid}). Assuming that all of the energy of the pebbles is deposited at the surface of the planet, this corresponds to a luminosity of
\begin{align}
\frac{G M_\mathrm{p} \dot{M}_{\rm{Hill}}}{R_\mathrm{p}} &= \Sigma_\mathrm{p} R_\mathrm{H}^2 v_{\rm{esc}}^2  \Omega\\
&\approx 2.7 \times 10^{28} \text{erg/s} \fid{M_\mathrm{p}}{M_\oplus}{4/3} \fid{r}{\text{AU}}{-1/2}
\end{align}
where we have again used our fiducial disk parameters, and assumed a density of $\rho_\mathrm{p} = 5.5 \, \text{g}/\text{cm}^3$ for the planet. Because of this extremely high accretion luminosity, planets undergoing pebble accretion will generally transport energy by convection through the entirety of their atmosphere. However, convection cannot transport an arbitrary amount of energy; for high enough luminosities convection will become inefficient, limiting the mass of the planet's atmosphere. 

In order to ensure that the atmospheric masses of planets undergoing pebble accretion were not too limited by pebble accretion, we numerically calculate steady state atmospheric masses following the methods of \citet{raf06}. The nebular parameters were calculated using the fiducial disk model discussed in Section \ref{fid}. We assume a simple power law opacity, $\kappa = \kappa_0 \left(T/T_0\right)$ 
, where $T_0$ is the temperature of the nebula at the given semi-major axis and $\kappa_0 = 0.1 \, \text{cm}^2 \text{g}^{-1}$. The temperature gradient $\nabla \equiv d \ln T / d \ln P$ was calculated using mixing length theory following Appendix D of \cite{raf06}. 

The results of this calculation are shown in Figure \ref{fig:m_atm} for $\alpha = 10^{-2}$. An analytic estimate of the mass of a fully convective atmosphere, $M_{\rm{atm}} \approx 4 \pi \rho_0 R_\mathrm{B}^3$, where $\rho_0$ is the nebular density, is also plotted. The solid lines are truncated on the left when $R_p > R_\mathrm{B}$. As can be seen in the figure, the atmospheric masses of these planets are generally very close to the fully convective value, with efficiency of convection only being important for the $M_\mathrm{p} = 10^{-1} \, M_\oplus$ planet past $a \sim 1 \, \text{au}$. 

A simple order of magnitude argument shows that the mass of a fully convective atmosphere is sufficient to repel the flow of nebular gas. Consider a core of mass such that $R_\mathrm{B} < R_\mathrm{H}$ 
with a fully convective atmosphere of mass $M_{\rm{atm}} \sim \rho_{\rm{neb}} R_\mathrm{B}^3$. The gas moves relative to the core's atmosphere with a velocity $v_{\rm{app}} \sim \max(\eta v_\mathrm{k}, \Omega R_\mathrm{B})$. In a time $\Delta t \sim R_\mathrm{B}/v_{\rm{app}}$ the core encounters a mass in gas of $M_{\rm{gas}} \sim \rho_{\rm{neb}} R_\mathrm{B}^2 v_{\rm{app}} \Delta t$, which therefore has kinetic energy $KE \sim \rho_{\rm{neb}} v_{\rm{app}}^2 R_\mathrm{B}^3$. The binding energy of the atmosphere is of order $E_{\rm{bind}} \sim G M_\mathrm{p} M_{\rm{atm}} / R_\mathrm{B}$. The ratio of these two quantities is therefore
\begin{align}
\frac{KE}{E_{\rm{bind}}} \sim \frac{\rho_{\rm{neb}} v_{\rm{app}}^2 R_\mathrm{B}^3}{\rho_{\rm{neb}} G M_\mathrm{p} R_\mathrm{B}^2} \sim \frac{v_{\rm{app}}^2 R_\mathrm{B}}{v_H^2 R_\mathrm{H}}
\end{align}
If $v_{\rm{app}} = \Omega R_\mathrm{B}$ then the quantity on the right is $<1$ since $R_\mathrm{B} < R_\mathrm{H}$ by assumption. Otherwise $v_{\rm{app}} = \eta v_\mathrm{k}$, in which case the quantity on the right is of order $c_\mathrm{s}^2/v_\mathrm{k}^2 = (H/r)^2 \ll 1$. In both cases the incoming kinetic energy of the gas is much less than the binding energy of the atmosphere, meaning the nebular gas will not ablate the stationary atmosphere. In particular, the ``recycling" effects identified by e.g. \citet{osk_2015} are unlikely to result in an unbound atmosphere during this phase of planetary growth.

\subsection{Pebble Isolation Mass} \label{peb_iso_mass}
In this section we discuss another candidate for limiting the growth of planets via pebble accretion, the ``pebble isolation mass,'' first identified by \citet{ljm_2014}. Once a planet reaches this mass scale, perturbations from the planet on the local gas disk raise pressure bumps in the disk that trap pebbles, preventing them from being accreted by the planet. From the results of their hydrodynamical simulations, \citeauthor{ljm_2014} give the pebble isolation mass as
\begin{align} \label{eq:M_peb_simp}
M_{\rm{iso}} = 20 M_\oplus \fid{H/r}{0.05}{3} \; .
\end{align}
Though it is not noted in \cite{LJ14}, this mass scale is similar in scale to the mass scale where $R_\mathrm{B} = R_\mathrm{H}$; specifically using the mass scale given in Equation \eqref{eq:M_th} without the factor $f$ and using the temperature profile used in \citet{ljm_2014} gives the semi-major axis scaling as in Equation \eqref{eq:M_peb_simp} with a prefactor of $\sim 23 M_\oplus$. 

\cite{bmj_2018} followed up on the work of \citep{ljm_2014} by exploring the variation of pebble isolation mass with the level of nebular turbulence and radial pressure gradient, and also accounted for how different pebble sizes are able to diffuse through the pressure bump raised by the planet. Their results confirm that the pebble isolation mass is of the scale of the thermal mass, with a variation of a factor of 2-3 as $\alpha$ is increased, and smaller effect from the radial pressure gradient. They also found that the mass of the planet must be increased an additional factor to block smaller particles; while the overall functional form of this increase is complicated, it is inversely proportional to the particle Stokes number.

Thus, in general the pebble isolation mass is of order the scale where $R_\mathrm{B} = R_\mathrm{H}$.
From our purely analytic arguments, i.e.  without the unknown order unity factor $f$, we expect the flow isolation mass to be of order this scale or smaller, (e.g. $\approx 30\%$ of this scale when $St_\mathrm{max} = 10^{-1}$) which would indicate that $M_\mathrm{flow} \lesssim M_\mathrm{peb}$ for small values of $St_\mathrm{max}$, with $M_\mathrm{flow} \sim M_\mathrm{peb}$ within a factor of 2-3 for $St_\mathrm{max} \sim 1$. A precise comparison is complicated by the dependence of the mass scales on the value of $f$, the value of $\alpha$, and to a lesser extent $\partial \ln P / \partial \ln r$. A more difficult to overcome complication stems from the dependence of $M_\mathrm{peb}$ on the \textit{smallest} Stokes numbers present: in contrast to the flow isolation mass, the pebble isolation mass more readily blocks large particles than small particles, meaning that the pebble isolation mass increases as the particle size that is required to be blocked is decreased. Because particles in protoplanetary disks do not exist at a single size, but instead have a distribution of sizes, in order to halt growth the planet must block not just the largest particles, but also sufficiently small particles such that the planet grows on timescales longer than the dissipation timescale of the protoplanetary disk. One could attempt to estimate this smallest particle size by assuming a size distribution for the small particles, and then calculating the smallest particle size below which the growth timescale for the core exceeded the lifetime of the gas disk. While we initially attempted this approach, we found that in many cases the calculated mass exceeded the regime where the analytic expressions of \citet{bmj_2018} hold. We therefore leave a detailed comparison between these two mass scales at high $St_\mathrm{max}$ to future work.

\section{Results} \label{results}
In this section we present values for the limiting mass that a growing planet can reach via pebble accretion by taking into account the flow isolation mass. We present results both fixed maximum Stokes number (Section \ref{fix_st}), and for a simple fragmentation limited model of particle size (Section \ref{frag_st}).  In Section \ref{m_star} we discuss how the flow isolation mass scales as a function of stellar mass.

\subsection{Limiting Planet Mass for Fixed $St_{\mathrm{max}}$} \label{fix_st}
In this section we give limits on planet mass as a function of the maximum Stokes number present in the disk. 
\begin{figure*} [htbp]
	\centering
		\includegraphics[width=7in]{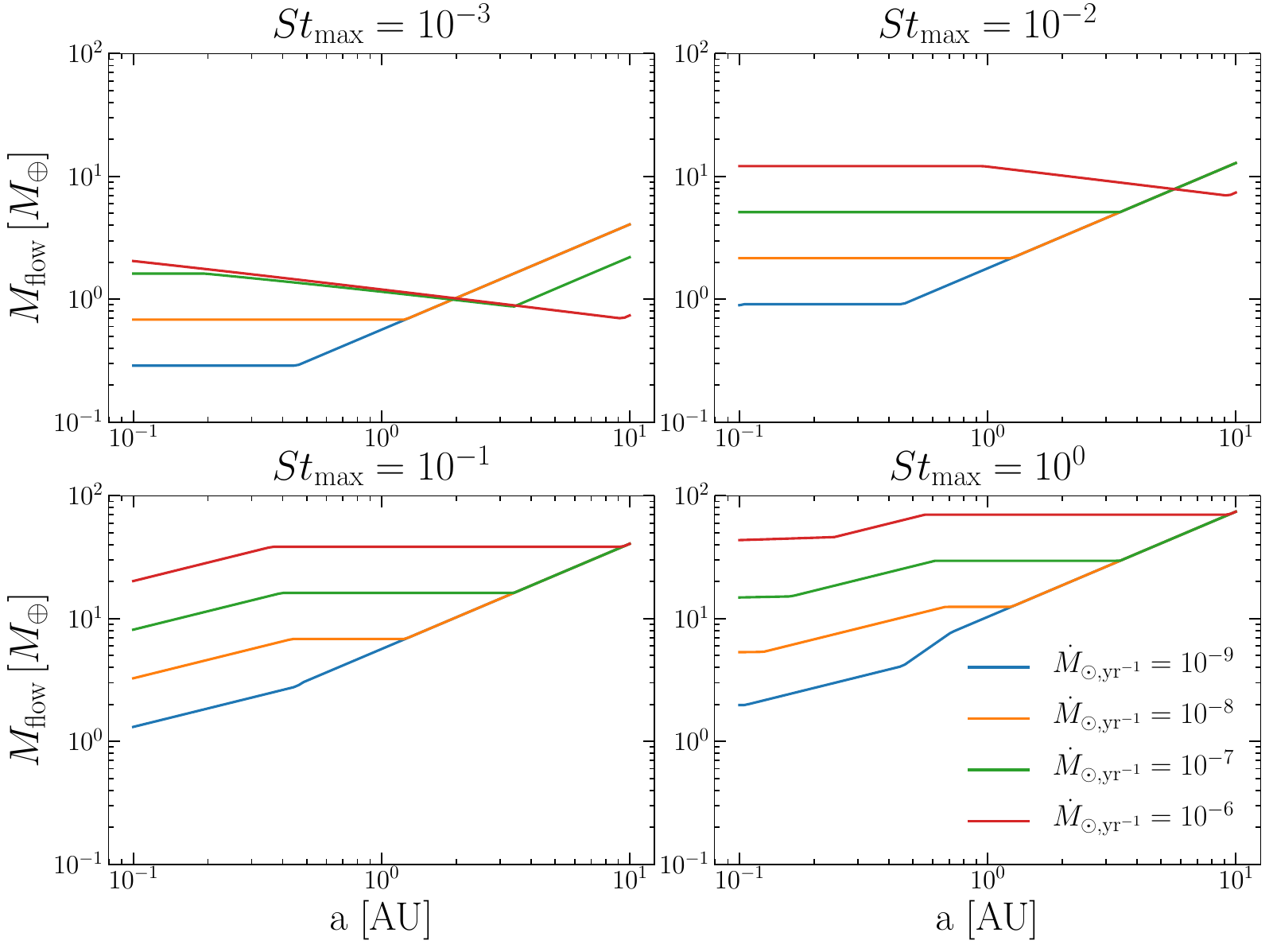}
		\caption{A plot of the maximal mass a planet accreting pebbles can reach as a function of semi-major axis, accretion rate, and maximum Stokes number present.}
\label{fig:m_limit_fixed_st}
\end{figure*}
Results for the flow isolation mass for fixed $St_{\rm{max}}$ are shown in Figure \ref{fig:m_limit_fixed_st}.

Several features are apparent in Figure \ref{fig:m_limit_fixed_st}. Firstly, the liming mass increases as a function of semi-major axis. However, the dependence is relatively shallow, particularly in the inner disk, where the mass can become independent of semi-major axis. In the inner region of the disk viscous heating dominates over irradiation; for planets at this mass scale the second of the three analytic expressions given in Equation \eqref{eq:m_flow_th} dominates, i.e., the flow isolation mass is given by
\begin{align}
    \frac{M_\mathrm{flow}}{M_\mathrm{th}} = \frac{f^{3/2}}{3} \sqrt{ St_{\mathrm{max}}}
\end{align}
Defining 
\begin{align}
    St_1 = \frac{St_\mathrm{max}}{10^{-1}}
\end{align}
then, scaled to our fiducial disk profile, this mass is given by
\begin{align} \label{eq:m_flow_sh_fid}
    M_\mathrm{flow} = \begin{dcases*}
     6.8 M_\oplus \, St_1^{1/2} r_{\mathrm{AU}}^{0} M_8^{3/8} M_{*,\odot}^{-1/8} \Sigma_{3000}^{3/8} \, & $r<r_\mathrm{vis-irr}$\\
    3.5 M_\oplus \, St_{1}^{1/2} r_{\mathrm{AU}}^{6/7} M_{*,\odot}^{-5/7} L_{*,\odot}^{3/7} \, & $r>r_\mathrm{vis-irr}$
    \end{dcases*}
\end{align}
i.e. the flow isolation mass is independent of semi-major axis in the inner disk, which is what causes the flattening of the lines seen in Figure \ref{fig:m_limit_fixed_st}.  Indeed, the scaling in Equation \eqref{eq:m_flow_sh_fid} may be complicated by several effects. When the Stokes number is low and the accretion rate is high (e.g., Figure \ref{fig:m_limit_fixed_st} top left, red line), the WISH radius can set the flow isolation mass rather than the shearing radius.  This causes $M_{\rm{flow}}$ to decrease with semi-major axis. Furthermore, close in to the star non-linear drag effects become important, causing $M_{\mathrm{flow}}$ to deviate from the simple scaling predicted by Equation \eqref{eq:m_flow_sh_fid}, as seen in the bottom two panels of Figure \ref{fig:m_limit_fixed_st}.

Finally, as can be seen in Figure \ref{fig:m_limit_fixed_st}, increasing the maximum Stokes number present in the disk increases the maximal mass planets can achieve. This is because larger particles can be captured at greater impact parameters, requiring the planet to reach higher masses before $R_\mathrm{B}$ overtakes $R_{\mathrm{stab}}$. In the next section, we consider how this maximal particle size might scale with semi-major axis. 

\subsection{Flow Isolation Mass for Fragmentation-Limited Pebbles} \label{frag_st}

\begin{figure} [htbp]
	\centering
        \includegraphics[width=1.1\linewidth]{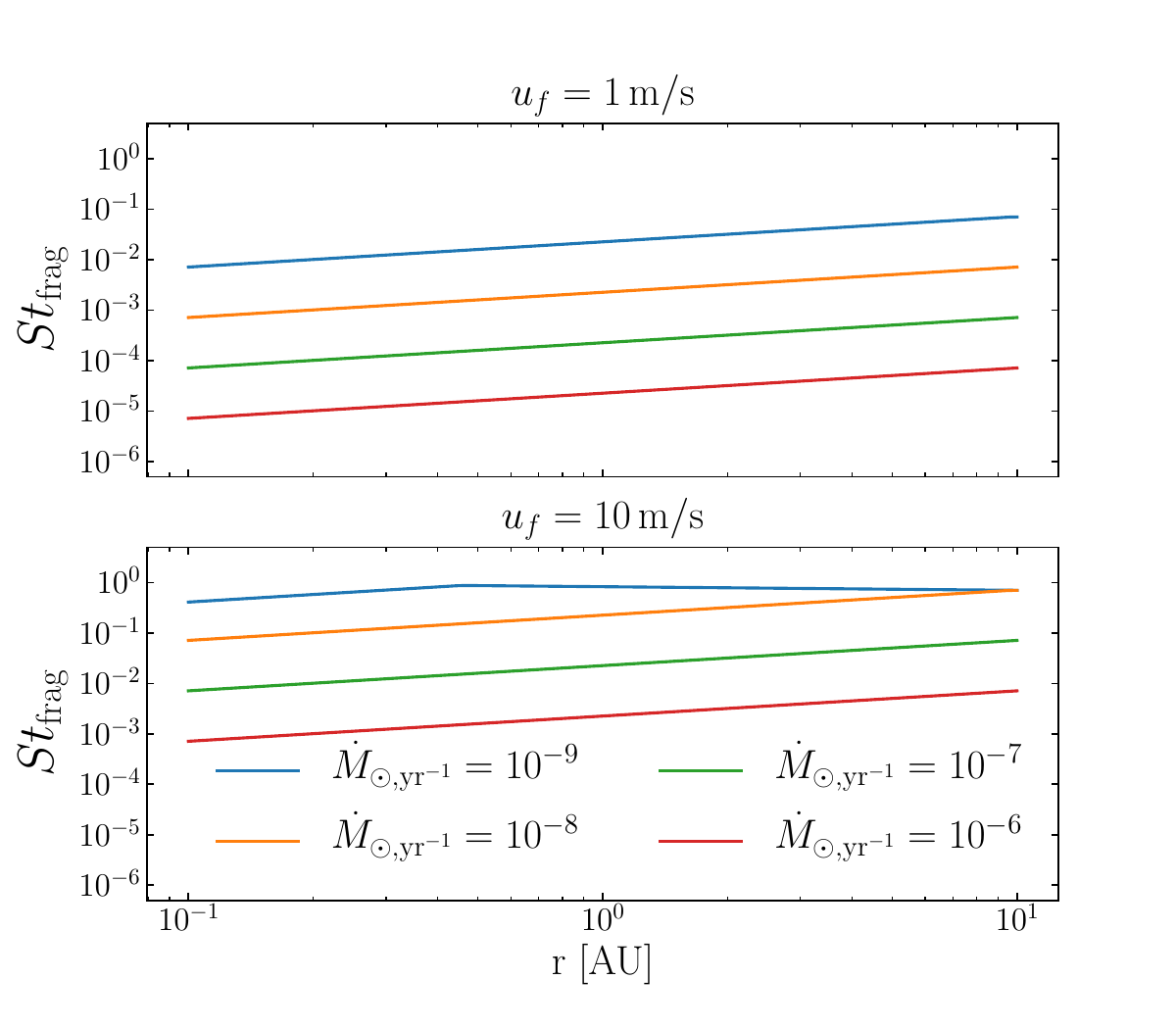}
		\caption{A plot of the maximal Stokes number pebbles can reach as a function of semi-major axis and particle fragmentation velocity. The maximal particle size at a given semi-major axis is given by Equation \eqref{eq:st_frag}.}
\label{fig:st_frag}
\end{figure}

\begin{figure*} [htbp]
	\centering
		\plotone{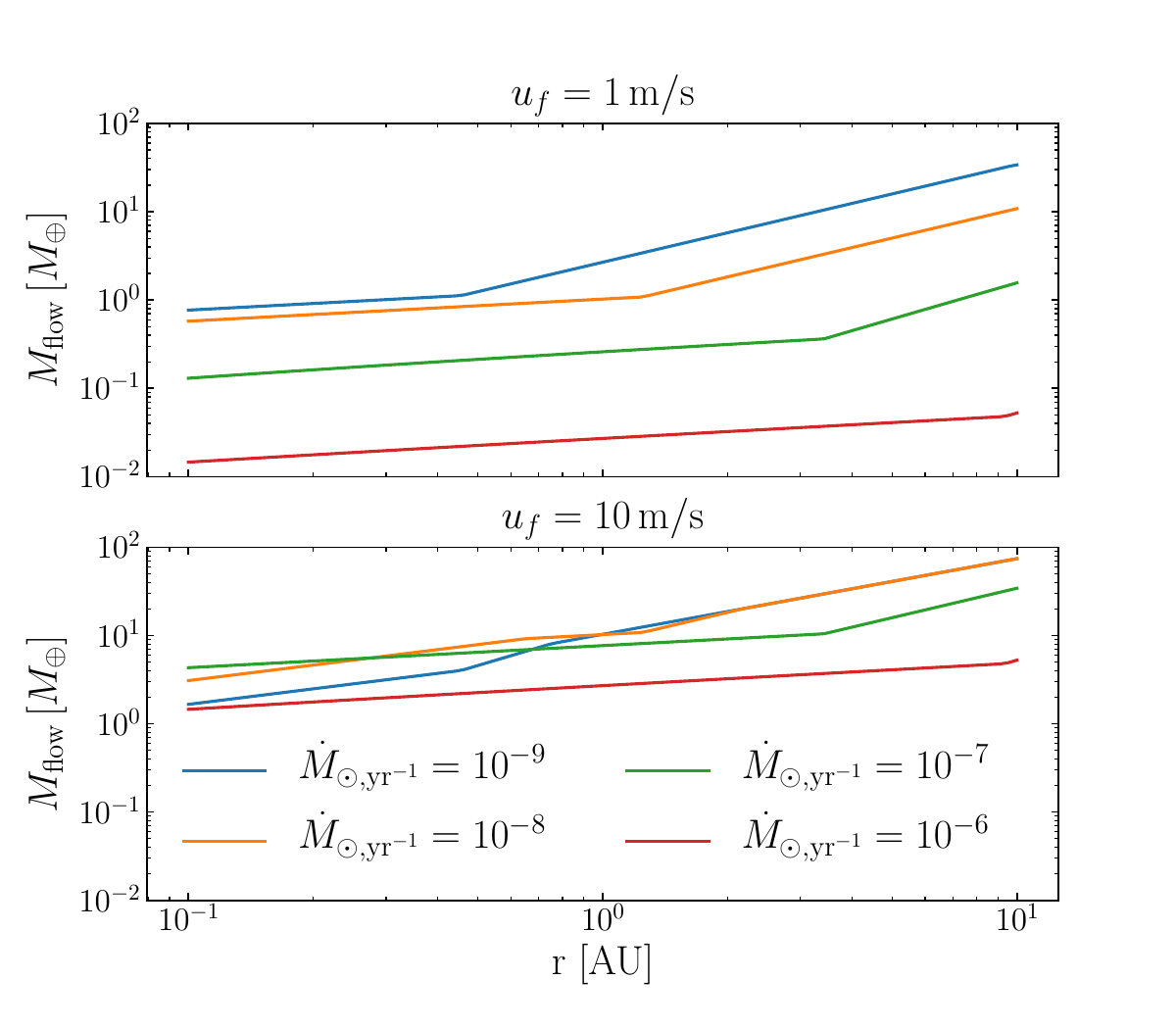}
        \epsscale{1.2}
		\caption{A plot of the maximal mass a planet accreting pebbles can reach as a function of semi-major axis and particle fragmentation velocity. The maximal particle size at a given semi-major axis is given by Equation \eqref{eq:st_frag}.}
\label{fig:m_limit_u_frag}
\end{figure*}

In the previous section we described the limiting planet mass as a function of Stokes number. There exist however, models for the maximal particle size present in the disk, which we can employ to remove the dependence on $St_{\rm{max}}$. In particular, in the inner regions of protoplanetary disks it is thought that fragmentation between particles limits the sizes that small bodies can reach, due to high collision velocities and frequent collisions. In this section we use a relatively simple model in which collision velocities above a threshold velocity $u_{\mathrm{frag}}$ result in fragmentation (e.g. \citealt{bdb_2009}). This would be expected if the binding energy of the particle scales as the particle's mass, which is an acceptable approximation for small solids held together by chemical bonds.  Lab experiments suggest that $u_{\mathrm{frag}}$ in the range 1-10 m/s may apply, though the (unknown) material properties of the colliding pebbles affect this number significantly \citep{sl_2009,blum_wurm_coll}.

If turbulent motions dominate the relative velocity between particles, then the relative velocity between two particles of with Stokes number $St$ is of order
\begin{align}
v_{\rm{coll}} = v_{\rm{gas},t} \sqrt{St}
\end{align}
assuming the particles have stopping times such that $t_\eta < t_\mathrm{s} < t_L$, where $t_\eta$ and $t_L$ are the turnover times of the smallest and largest scale eddies respectively \citep{oc07}. This leads to a maximum Stokes number of
\begin{align} \label{eq:st_frag_turb}
St_{\rm{max}} = \frac{u_{\rm{frag}}^2}{\alpha c_\mathrm{s}^2}
\end{align}
\cite{bdb_2009}. In what follows, we also consider collisions stemming from the laminar gas velocity. For particles with $St < 1$, the particle's laminar velocity relative to Keplerian is well approximated by $v_\ell = 2 \eta v_\mathrm{k} St$, leading to a relative velocity of
\begin{align}
v_{\rm{rel},\ell} = 2 \eta v_\mathrm{k} \left(St_1 - St_2 \right) \sim \eta v_\mathrm{k} St_1
\end{align}
where $St_1$ and $St_2$ are the Stokes numbers of the larger and smaller particles, respectively. This leads to a maximum Stokes number of roughly
\begin{align} \label{eq:st_frag_lam}
St_{\rm{max}} = \frac{u_{\rm{frag}}}{\eta v_\mathrm{k}} \; .
\end{align}
Combining Equations \eqref{eq:st_frag_turb} and \eqref{eq:st_frag_lam} gives
\begin{align} \label{eq:st_frag}
    St_{\mathrm{max}} = \min \left(\frac{u_{\rm{frag}}^2}{\alpha c_\mathrm{s}^2}, \frac{u_{\rm{frag}}}{\eta v_\mathrm{k}} \right)
\end{align}
The maximum Stokes number for fragmentation velocities of $u_{\rm{frag}} = 1 \, \text{m/s}$ and $10 \, \text{m/s}$ are shown in Figure \ref{fig:st_frag}.

In Figure \ref{fig:m_limit_u_frag} we plot the value of the flow isolation mass for fragmentation velocities of $u_{\rm{frag}} = 1 \, \text{m/s}$ and $10 \, \text{m/s}$. For a fragmentation velocity of $u_{\rm{frag}} = 1 \, \rm{m/s}$ (upper panel), only the colder disks, i.e. those with lower $\dot{M}$, are able to produce super-Earth masses. For $u_{\rm{frag}} = 10 \, \rm{m/s}$ (lower panel), however, the mass scale is much less sensitive to the temperature. This is because there are two competing effects that tend to cancel one another out as the temperature is increased: higher temperatures increase the thermal mass, increasing the flow isolation mass as well. However, higher temperatures lead to larger collision velocities between particles, which decreases the maximum Stokes number and correspondingly lowers the flow isolation mass.

We comment that a given protoplanetary disk likely evolves at different accretion rates during its lifetime. This implies that  the final mass a planet reaches depends on when the initial protoplanet forms, as was also identified by \citet{bij_2019}. 

We emphasize that because both pebble accretion timescales for growing cores and collisional growth/destruction destruction timescales for source pebbles are very fast, particularly in the inner disk,  planets are likely able to reach the maximum flow isolation masses shown in Figure \ref{fig:m_limit_u_frag}.

\subsection{Variation with Stellar Mass} \label{m_star}

In Equation \eqref{eq:m_flow_sh_fid}, we gave the scaling of the flow isolation mass with fiducial disk parameters.  However, two of these quantities, $M_8$ and $\Sigma_{3000}$ likely scale with stellar mass. A number of observational works point to $\dot{M}$ scaling with $M_*^2$ (e.g. \citealt{ntr_2006}, \citealt{anm_2014}), and recent work points to a linear or steeper than linear scaling of disk mass with stellar mass (\citealt{ark_2013}, \citealt{pth_2016}). If we neglect variation in the outer disk radius, then this implies that the surface density also scales linearly with stellar mass. Inserting these scalings into the inner, viscously heated regime of Equation \eqref{eq:m_flow_sh_fid} (and assuming that our fiducial value of $
\Sigma_0 = 3000 \, \mathrm{g} \, \mathrm{cm}^{-2}$ applies for $M_{*,\odot} = 1$) gives
\begin{align} \label{eq:m_flow_stellar}
    M_{\mathrm{flow}} = 6.8 M_\oplus \, St_{1}^{1/2} M_{*,\odot}^1
\end{align}
i.e. the flow isolation mass scales approximately linearly with the host star mass.

\section{Comparison to Observations of Close in Planets} \label{obs}
In this section we compare predictions made if planet mass is limited by the flow isolation mass to trends identified in the population of close in planets. We point out that these trends are readily explained if planet mass is limited by flow isolation. For a discussion of super-Earth observations in the context of pebble isolation, see \citet{bij_2019}.

\subsection{\citet{wmp_2018} and \citet{mwl_2017}}

\cite{wmp_2018} investigated the characteristics of the multi-planet systems found in the \textit{Kepler} sample. These authors found a correlation between the sizes of planets within a given multi-planet system, i.e. planets in the same system are likely to be similar in size. \citet{mwl_2017} further showed through analysis of the Kepler planets that also have masses measured through transit timing variations that this similarity applies to mass as well as radius. Note however, that some authors have attributed this effect to detection bias \citep{z_2019}.

Similarity between sizes of planets emerges naturally if planet mass is limited by flow isolation. Looking at Equation \eqref{eq:m_flow_sh_fid}, we see that in the inner regions of protoplanetary disks the flow isolation mass is roughly independent of semi-major axis, which stems from viscous heating dominating the temperature structure in this region. Thus, if flow isolation limits planetary growth, super-Earths in the same system would be similar in mass.  Excluding atmospheric loss effects, they would also be similar in size.

\subsection{\cite{w_2019}}
Using updated radius values for planets found from the Kepler mission in concert with Gaia DR2 stellar radii, \cite{w_2019} explored the effects of photoevaporation in sculpting the observed super-Earth population. \citeauthor{w_2019} found that this population could be explained as stemming from a single characteristic mass scale, of roughly $M_\mathrm{p} \sim 8 M_\oplus$. Furthermore, \citeauthor{w_2019} demonstrated that this mass scale varies with stellar mass and radius with a power law indicies in the range 
\begin{align}
\label{eqn:wuscale}
    M_\mathrm{p} = 8 M_\oplus M_*^{0.95-1.4} r_{\mathrm{AU}}^{0-0.5}
\end{align}
Note that this mass scale refers to the bare core mass of these planets; planets that do not undergo photoevaporation will accrete some amount of nebular gas, changing their observed radius (and, to a lesser extent, mass).

Comparison between Equations \eqref{eq:m_flow_stellar} and \eqref{eqn:wuscale} shows that the scaling of this characteristic mass scale is exactly what we would expect if pebble accretion fuels the growth of these planets, only to be shut off by flow isolation. We note that \citet{w_2019} argues the characteristic mass scale identified in that work could be the thermal mass, whereas we have argued that this mass scale could be the flow isolation mass, which is generally less than or equal to the thermal mass. This difference stems from how the temperature profile in the inner regions of the protoplanetary disk and the scaling of various disk parameters with stellar mass are modeled.

\subsection{\citet{zw_2018} \& \citet{bkl_2019}}
Using previously published planetary systems, \citet{zw_2018} calculated the correlation between systems with ``cold" Jupiters and inner super-Earths. They found that 90\% of systems that host an outer cold Jupiter contain inner super-Earths. \citet{bkl_2019} further investigated the occurrence rate of such outer gas giant companions in systems that contain super-Earths by taking radial velocity data on systems containing super-Earths and looking for trends in the radial velocity signals. They found an occurrence rate of 39\% $\pm$ 7\% for planets 0.5-20 $M_{\mathrm{jup}}$ at 1-20 au, and also demonstrated that systems that host super-Earths are more likely to contain an outer gas giant planet. 

This effect would follow naturally for systems of super-Earths where the mass of the planets is limited by flow isolation. In such systems, solid surface densities and pebble sizes were clearly conducive to formation of planets via pebble accretion in the inner disk. At larger semi-major axes, the disk temperature is set by passive irradiation instead of viscous heating, indicating a weaker scaling of temperature with semi-major axis. This weaker scaling leads to larger values of thermal mass in the outer disk, and correspondingly larger flow isolation masses. Thus, in the outer regions of these disks the flow isolation mass can reach values large enough to trigger runaway gas accretion, allowing gas giants to form at larger semi-major axes (c.f. Figures \ref{fig:m_limit_fixed_st} and \ref{fig:m_limit_u_frag}, upward trends at righthand sides of plots).
Therefore, in systems which produced inner super-Earths via flow isolation, we would expect outer gas giants to be more likely, in line with the results of \citet{zw_2018} and \citet{bkl_2019}.  We note that at very large semi-major axes, drift limits the sizes of available pebbles \citep[e.g.,][]{pmp_2019}, meaning that the trend toward larger flow isolation masses will likely reverse at large separations. We also point out that this correlation between inner super-Earth and outer gas giants is not unique to the flow isolation mass, but is a natural prediction of theories where a limiting mass scale increases in the outer disk, as is true for the pebble isolation mass, e.g. \citet{bab_2018}, \citet{bij_2019}, or the local isolation mass used in classic models of the solar system (e.g. \citealt{l_1993}).

\section{Summary and Conclusions} \label{conclusions}
We discussed how pebble accretion timescales vary as a function of core mass, and pointed out that at super-Earth masses growth timescales for pebble accretion are extremely rapid for a large range of pebble sizes. These rapid growth rates make it difficult to form super-Earths via pebble accretion unless something halts growth once planets reach this mass scale. 

We further demonstrated that modification of the gas flow pattern by the planet's atmosphere limits accretion of the smallest pebble sizes. The Stokes number of the smallest pebble size a planet can accrete can be determined by finding the size for which the maximal impact parameter for accretion, $R_{\mathrm{stab}}$, is equal to the scale of the core's atmosphere, $R_\mathrm{B}$. If the solids present in the protoplanetary disk are limited to sizes smaller than a maximum size, then this process naturally predicts that growth of planet will cease once the minimum-sized particles a planet can accrete is larger than the maximal size present in the disk. For a reasonable fiducial disk profile and particle sizes, we showed that the resulting mass scale where growth ceases is around super-Earth masses.

Furthermore, we showed that several trends present in the demographics of the super-Earth population follow naturally if the masses of these planets are limited by flow isolation: super-Earths in the same system would be correlated in mass and radius, as reported by \citet{wmp_2018}, due to the shallow scaling of the flow isolation mass with semi-major axis in the inner disk. We would also expect a characteristic mass scale, i.e. the flow isolation mass, to be present in the super-Earth population, and to scale approximately linearly with stellar mass and weakly with semi-major axis, as reported by \citet{w_2019}. Finally, we would expect systems that have inner super-Earths to be more likely to host an outer gas giant, as the the flow isolation mass is larger at these larger orbital separations, a trend which was detected by \citet{zw_2018} and \citet{bkl_2019}.

While the trends in the super-Earth population seem consistent with being limited to the local flow isolation mass, there remain other regimes where the importance of the flow isolation mass could be tested, particularly in contrast with the pebble isolation mass. One such regime would be planet formation in the outer regions of protoplanetary disks -- in these regions maximal Stokes numbers are likely set by drift (e.g. \citealt{bke_2012}), which leads to maximal Stokes number of $St \sim 10^{-1}-10^{-2}$. On the other hand, the thermal mass is quite large in the outer disk, as the aspect ratio of the disk generally increases as a function of semi-major axis. Thus, in this region we would expect the predictions of flow isolation and pebble isolation to be quite different, with flow isolation predicting substantially lower planetary masses.

\vspace{1mm}

\noindent We thank Eugene Chiang and Hagai Perets for useful discussions. We thank the anonymous referee for their comments which improved the quality of the manuscript. MMR and RMC acknowledge support from NSF CAREER grant number AST-1555385.

\bibliographystyle{yahapj}
\bibliography{gas_bib}
\end{document}